\documentclass[final,5p,times,twocolumn,authoryear]{elsarticle}
\pdfoutput=1
\usepackage{graphicx}
\usepackage{amssymb}
\usepackage{listings}
\usepackage{nameref}
\usepackage[draft=true]{minted}
\usepackage{csquotes}
\usepackage{rotating}
\usepackage{hyperref}

\journal{Astronomy and Computing}

\begin{document}

\begin{frontmatter}

\title{From FATS to feets: Further improvements to an astronomical 
       feature extraction tool based on machine learning}

  \author[iate,fceia]{J. B. Cabral}\corref{mycorrespondingauthor}
      \cortext[mycorrespondingauthor]{Corresponding author}
      \ead{jbcabral@oac.unc.edu.ar}
  \author[iate,famaf]{B. S\'anchez}
  \author[iate]{F. Ramos}
  \author[iate]{S. Gurovich}
  \author[cifasis]{P. Granitto}
  \author[escience]{J. Vanderplas}
  
\address[iate]{
   Instituto De Astronom\'ia Te\'orica y Experimental -
   Observatorio Astron\'omico C\'ordoba (IATE--OAC--UNC--CONICET),
   Laprida 854, X5000BGR, C\'ordoba, Argentina}
\address[fceia]{
   Facultad de Ciencias Exactas, Ingenier\'{i}a y Agrimensura, UNR,
   Pellegrini 250 - S2000BTP, Rosario, Argentina}
\address[cifasis]{
   Centro Internacional Franco Argentino de Ciencias de la
   Informaci\'on y de Sistemas (CIFASIS, CONICET--UNR),
   Ocampo y Esmeralda, S2000EZP,
   Rosario, Argentina}
\address[escience]{
 	University of Washington eScience Institute,
    Campus Box 351570,
    University of Washington,
	Seattle, WA 98195-1570}
\address[famaf]{
	Facultad de Matem\'atica, Astronom\'{\i}a y F\'{\i}sica
    Universidad Nacional de C\'ordoba (FaMAF--UNC)
	Bvd. Medina Allende s/n, Ciudad Universitaria,
    X5000BGR, C\'ordoba, Argentina 
}

\begin{abstract}
Machine learning algorithms are highly useful for the classification of time series 
data in astronomy in this era of 
peta-scale public survey data releases.
These methods can facilitate the discovery of new unknown events in most astrophysical 
areas, as well as improving the analysis of samples of known phenomena.
Machine learning algorithms use features extracted from collected data as input predictive variables.
A public tool called Feature Analysis for Time Series (FATS) has proved 
an excellent workhorse for feature extraction, particularly light curve 
classification for variable objects. 
In this study, we present a major improvement to FATS, which 
corrects inconvenient design choices, minor details, and
documentation for the re-engineering process.
This improvement comprises a new Python package  called \textit{feets}, which is important 
for future code-refactoring for astronomical software tools.
\end{abstract}


\end{frontmatter}
% =============================================================================
% SECTION INTRO
% =============================================================================
\section{Introduction.}
\label{section:intro}
Machine learning (ML) has proved to be an important tool for data 
analysis in astronomy. Numerous projects such as AstroML \citep{vanderplas_astroml:_2014}, -
UPSILoN \citep{kim_package_2016}, MeSsI \citep{de_los_rios_messi_2016}, and Feature Analysis for Time Series (FATS)\citep{nun2015fats} have been developed to 
help astronomers use the ML approach.
Languages such as R \footnote{\url{https://www.r-project.org/}} and Python\footnote{\url{https://www.python.org/}} provide massive collections of ready-to-use
packages for complex data analysis and a large number of ML tools. 
However, if we conduct in-depth analysis of these systems,
it is apparent that some projects lack adequate software engineering design \citep{cowling_first_1998}, 
and thus they are often difficult to test, maintain, and extend, 
and some of these solutions may also under-perform in terms of speed and memory. 
Moreover, many projects run in critical environments and process large 
volumes of data over short timescales, and thus they can be inefficient if they are
not optimized correctly. 
Thus, in simple terms, many scientific systems have a ``bad code smell'' \citep{tufano_when_2015}.

According to Fowler, code smell is a: 
\begin{quotation}
 \dots a surface indication that usually 
corresponds to a deeper problem in the software; and we currently 
know that \textit{most of times code artifacts are affected by so-called 
``bad smells'' since their creation}\dots  \citep{fowler_martin_codesmell_2006}. 
\end{quotation}
In these cases, the most pragmatic solution is to replace most of the 
code for a superior implementation, but to avoid any functional changes.
This type of process is called \textit{code refactoring} \citep{fowler_refactoring:_1999}.
 
In this study, we employed a code refactoring process to 
provide a more robust time-series feature extraction library based on 
the \textit{FATS} project \citep{nun2015fats}. 
The current version of FATS (1.3.6)\footnote{\url{https://pypi.org/project/FATS}}
is written entirely in Python and 
based on the numerical libraries comprising Numpy \citep{walt_numpy_2011}, 
Scipy \citep{jones_scipy:_2014}, and StatsModels \citep{seabold_statsmodels:_2010}.
FATS can extract up to 64 features from time series data inputs,
and it also includes pre-processing functions and tools for importing data from
the MACHO survey \citep{cook_variable_1995}. 
The project tutorial for feature extraction\footnote{\url{http://isadoranun.github.io/tsfeat/FeaturesDocumentation.html}} is instructive,
but it lacks internal documentation, which is crucial when adding features to this otherwise excellent tool.
In particular, we identified several limitations
when we attempted to use FATS for classifying periodic variable 
stars from large data sets, such as the VVV/VVV(x) \citep{catelan_vista_2011, minniti_mapping_2018} 
projects.
Thus, we engineered an upgrade, which was carefully designed to build on the strengths of FATS by re-utilizing as much code 
and documentation as possible.

The remainder of this study is organized as follows.
In \autoref{section:features}, we explain the feature extraction formalism and the
relevance of automatic classification. In
\autoref{section:FATS}, we consider the original project to identify 
advantages and weaknesses. 
The theoretical background of the re-engineering process
and the selected actions required to implement \textit{feets} are described in 
\autoref{section:reengineering}. In \autoref{section:results}, we provide 
detailed explanations of the internal details that make 
\textit{feets} a better choice for extracting time series features.
Finally, in \autoref{section:conclusions}, we give our conclusions and 
suggestions for future improvements to the FATS project.

% =============================================================================
% SECTION FEATURES
% =============================================================================
\section{Feature engineering and ML}
\label{section:features}
ML algorithms can be applied to large 
volumes of data in order to improve their performance at a given task \citep{samuel_studies_1959}.
These tasks may include classification, regression, optimization, or clustering, which are
the most common ML applications \citep{michalski_machine_2013}.
The data employed may originate from a wide range of sources, but 
individual observations are always represented by a set of values called features.
The process conducted for defining and extracting these features is called 
feature engineering. 
The highly specific nature of the feature engineering process makes it expensive, difficult, 
and time consuming, where it requires expertise in the area of application, and it is 
also the most critical step in a 
ML project \citep{ng_machine_2013}.

% =============================================================================
% SECTION FATS
% =============================================================================
\section{FATS }
\label{section:FATS}
The FATS tool is used to extract
characteristics from time series data. 
In particular, the FATS project aims to
standardize the feature extraction process for astronomical
light curves. FATS is built on top of the Python scientific stack (Numpy and
Scipy) and it also uses StatsModels for additional statistical analysis.
\subsection{Simple example}
A sample 
from the original documentation is shown below to clearly illustrate the different components of FATS.
\begin{quotation}
The library receives as input the time series data and returns as output 
an array with the calculated features. Depending on the available input 
the user can calculate different features. 
For example, if the user has only the vectors magnitude and time, 
just the features that need this data will be able to be computed.
\end{quotation}
Obviously the number of data points, data cadence, etc. will also 
influence the number of possible features. 
\begin{quotation}
In order to calculate all the possible features the following vectors 
(also termed as raw data) are needed per light curve:
\begin{itemize}
\item magnitude
\item time
\item error
\item magnitude2
\item aligned\_magnitude
\item aligned\_magnitude2
\item aligned\_time
\item aligned\_error
\item aligned\_error2
\end{itemize}
where \textit{2} refers to a different observation band.
It should be noted that the magnitude vector is the only 
input that is strictly required by the library because it is necessary 
for calculating all the features. The remaining vectors are 
optional because they are only needed by some features. 
Thus, if the user does not have these additional data or 
time series other than light curves are being analyzed, it is still possible 
to calculate some of the features (\dots)
\end{quotation}

We illustrate this point with the following Python code, 
which calculates features for a randomly generated light curve. 
Only magnitude and time data are used, so the smallest feature set will be obtained. This code is also based on the FATS tutorial.
\begin{minted}[]{pycon}
>>> import numpy as np
>>> import FATS

# randomly generate the data
>>> magnitude_ex = np.random.rand(30)
>>> time_ex = np.arange(0, 30)

# create the light curve array with the same 
# order as the previous list
>>> lc_example = np.array([magnitude_ex, time_ex])

# create the feature space (this object serves 
# as an entry point for extracting all the features) 
# by specifying the available data
>>> fs = FATS.FeatureSpace(
...     Data=['magnitude','time'])
Warning: the feature Beyond1Std could not be 
   calculated because 
  ['magnitude', 'error'] are needed.
  ...
Warning: the feature CAR_mean could not be 
   calculated because 
  ['magnitude', 'time', 'error'] are needed.

# calculate the features of the light curve
>>> fs.calculateFeature(
...   lc_example).result("dict")
{'Amplitude': 0.46422830004583993,
 'AndersonDarling': 0.69055170152838952,
 'Autocor_length': 1.0,
 'Con': 0.0,
 'Eta_e': 1.2660816816234817,
 ...
 'Period_fit': 0.99987790236635843,
 'Psi_CS': 0.25119898781877714,
 'Psi_eta': 1.4069202501024807,
 'Q31': 0.47029269745864666,
 'Rcs': 0.21319202165853643,
 'Skew': 0.19709543035122007,
 'SmallKurtosis': -0.93310208660425609,
 'Std': 0.28904604267318268}
\end{minted}

\subsection{Functionalities}

In addition to the previous example,
FATS provides the following functionalities.
\begin{description}
\item [Features Extraction Framework:]
FATS includes a simple tool for creating your own feature extractor.
Every feature extractor is a python class inside the 
\texttt{FATS.FeatureFunciontLib} module. 
This class must contain object \textit{attributes} that define the 
\texttt{data} required for the light curve
and a \texttt{fit()} \textit{method} used for calculating the feature.
The following code is an example of a class that returns the number of observations
in a time series.
    
\begin{minted}[]{python}
class Count(Base):
  def __init__(self):
    # this attribute sets the 
    # data required to calculate the feature
    self.Data = ['time']

  def fit(self, data):
    # the extractor retrieves the 
    # time data from the light curve
    time = data[1]  
     
    # finally, the extractor simply 
    # returns the number of 
    # observations
    return len(time)
\end{minted}
Next, the extractor can be used as follows:
\begin{minted}[]{pycon}
# create the feature space 
# specify the required feature data
>>> fs = FATS.FeatureSpace(
...   featureList=["Count"])
  
# calculate the features of the light curve
>>> fs.calculateFeature(
...   lc_example).result("dict")
{'Count': 30}
\end{minted}

\item [Time Series Preprocessing:]
Two functions are integrated for light curve data pre-processing, where 
the first is a time series error parameter based on a sigma-clipping 
algorithm (as implemented in the \texttt{FATS.Preprocess\_LC} class) 
and the second is a class called 
\texttt{FATS.Align\_LC} for aligning two time series.
The documentation suggests that these functionalities are used before 
any features are extracted.
\item [MACHO-survey Light-Curve Parser:]
The \texttt{FATS.ReadLC\_MACHO} class retrieves the magnitude, time, and error 
from a given MACHO-id object (the id is assigned in the MACHO survey). 
This implementation does not search any of the data from the MACHO survey, 
but instead the user is responsible for downloading the light curve to the current 
working directory.
\end{description}

\subsection{Advantages and disadvantages of FATS}
\label{section:light_shadows}
From a software engineering perspective, many design decisions 
in the FATS project are good implementations, whereas others are inadequate. 
In the following, the "good" design decisions are described in short
paragraphs, before focusing on the problematic
design decisions. 

\subsubsection{Good design choices}
\label{good_design_choices}
A good design choice in FATS is separating the 
API\footnote{Abstract programming interface: The collection of functions, classes, 
and objects that the programmer can use the library.} into two parts as follows.
\begin{enumerate}
\item Extraction is configured in the \texttt{FATS\.FeatureSpace} class.
\item The extraction of configured features is implemented as a class hierarchy 
      inside the \texttt{FATS\.FeatureFunciontLib} module.
\end{enumerate}
This split between the functionality for analysis (\texttt{FATS\.FeatureSpace}) and
the creation of the feature extraction framework simplifies the operation of the system,
but allows the possibility of implementing complex feature extractors.
In addition, as noted above, preprocessing and MACHO light curve manipulation are not
associated directly with the core functionality of the library, but instead they are provided as an 
additional option in the project.

\subsubsection{Criticisms}
\label{section:criticism}
We consider several weak aspects of the current FATS implementation where 
the code is analyzed in depth.
Some problems are simple style errors. 
Others are related to the development process and 
design bugs, which may cause errors and limitations during feature extraction.

The complete experiment that forms the basis of many of our criticisms can be 
found at: 
\url{https://github.com/carpyncho/feets\_paper/blob/master/reports/FATS\_tests.ipynb}

\begin{description}
\item [Style and Maintainability] 
  Python has a strict coding style defined
  in Python Enhancement Proposal 8 (PEP-8) \footnote{\url{https://www.python.org/dev/peps/pep-0008/}}.
  This document defines guidelines for making code easy to understand by any
  Python developer. When a project follows these guidelines, as well as others such as
  PEP-20 \footnote{\url{https://www.python.org/dev/peps/pep-0020/}} 
  (related to the philosophy behind python design), the Python community refers to 
  the code as "Pythonic" (easy to understand and maintain). 
  PEP-8 errors can be checked easily with several tools such as 
  flake8\footnote{\url{http://flake8.pycqa.org}} 
  and pylint \footnote{\url{https://www.pylint.org/}}, so avoiding
  style errors is a straightforward task. 
  
  FATS does not adhere to the recommendations and $828$
  style errors were found in $1249$ lines of code in a recent executed count 
  (i.e., $66\%$ of the lines contained errors if we assume a
  uniform distribution of errors).

\item [Global Configurations]
  The FATS documentation states the following\footnote{\url{http://isadoranun.github.io/tsfeat/FeaturesDocumentation.html}}.
  \begin{quotation}
  \textbf{Note:} Some features depend on other features and consequently 
  must be computed together. For instance, \texttt{Period\_fit} returns the 
  false alarm probability of the estimated period. 
  Thus, it is also necessary to calculate the period \texttt{PeriodLS}.
  \end{quotation}

  These dependencies are implemented as global variables. 
  For example, we can check this with the 
  \texttt{StructureFunction\_index\_21} class from the 
  \texttt{FATS.FeatureFunctionLib} module.
  
\begin{minted}[linenos]{python}
class StructureFunction_index_21(Base):

  def __init__(self):
    self.Data = ['magnitude', 'time']
  
  def fit(self, data):
      magnitude = data[0]
      time = data[1]

      global m_21
      global m_31
      global m_32

      # more code here

      m_21, b_21 = np.polyfit(sf1_log, sf2_log, 1)
      m_31, b_31 = np.polyfit(sf1_log, sf3_log, 1)
      m_32, b_32 = np.polyfit(sf2_log, sf3_log, 1)

      return m_21
  \end{minted}
  
According to this example, the \texttt{m\_21}, 
\texttt{m\_31}, and \texttt{m\_32} (lines 16--18) variables are calculated and stored in the
\texttt{global} environment/module level (lines 10--12) but only the 
\texttt{m\_21} value is returned (line 20). 

In addition, if we check the \texttt{StructureFunction\_index\_31} class
in the same module:
\begin{minted}[linenos]{python}
class StructureFunction_index_31(Base):
  
  def __init__(self):
    self.Data = ['magnitude', 'time']
  
  def fit(self, data):
     try:
       return m_31
     except:
       print ("error: please run "
              "StructureFunction_index_21 "
              "first...")
\end{minted}
we can see that this class is only used to retrieve the 
\texttt{m\_31} variable from the global 
environment (line 31), or to print an error to the console (line 10).

This design choice creates a bug, which can be  
exploited to retrieve incorrect values according to the following simple
procedure.
\begin{enumerate}
\item First, import the modules and create two synthetic light curves: 
	\texttt{normal\_lc} (the magnitudes are generated from a Gaussian distribution of values) 
	and \texttt{uniform\_lc} (the magnitudes are generated from a uniform value distribution).
\begin{minted}[]{pycon}
>>> import numpy as np
>>> import FATS

>>> mag = np.random.normal(size=10000)
>>> time = np.arange(10000)
>>> normal_lc = [mag, time]

>>> mag2 = np.random.uniform(size=10000)
>>> time2 = np.arange(10000)
>>> uniform_lc = [mag2, time2]
\end{minted}

\item Second, create a feature space from which to extract the main built-in features comprising 
	\texttt{StructureFunction\_index\_21} and \texttt{StructureFunction\_index\_31} from the 
	normal light curve (\texttt{normal\_lc}).
\begin{minted}[]{pycon}
>>> fs_normal = FATS.FeatureSpace(
... featureList=[
...   'StructureFunction_index_21',
...   'StructureFunction_index_31'])

# extract the features
>>> fs.calculateFeature(normal_lc)
>>> result = fs.result(method='dict')

# print the results
>>> print "Normal LC:"
>>> for f, v in result.items():
...   f = f.split("_", 1)[-1]
...   print "  {} = {}".format(f, v)
Normal LC:
  index_21 = 1.97547953389
  index_31 = 3.05091739197
\end{minted}
\item Create a \textbf{new} feature space and try to extract only the 
	\texttt{StructureFunction\_index\_31} feature from the uniform light curve.
    (\texttt{uniform\_lc}). The value obtained should be the same as that for the 
	\texttt{StructureFunction\_index\_31} of the normal light curve.

\begin{minted}[]{pycon}
>>> fs2 = FATS.FeatureSpace(
... featureList=[
...   'StructureFunction_index_31'])

# extract the features
>>> fs2.calculateFeature(uniform_lc)
>>> result = fs2.result(method='dict')

>>> print "Bad Uniform LC:"
>>> for f, v in result.items():
...   f = f.split("_", 1)[-1]
...   print "  {} = {}".format(f, v)
Bads Uniform LC:
  index_31 = 3.05091739197
\end{minted}
\end{enumerate}

\item The only way to avoid this problem is to calculate 
	\texttt{StructureFunction\_index\_21} by default for all of the input light curves.

\begin{minted}[]{pycon}
>>> fs3 = FATS.FeatureSpace(
... featureList=[
...   'StructureFunction_index_21',
...   'StructureFunction_index_31'])

# extract the features
>>> fs3.calculateFeature(uniform_lc)
>>> result = fs3.result(method='dict')

>>>  print "Uniform LC:"
>>> for f, v in result.items():
...   f = f.split("_", 1)[-1]
...    print "  {} = {}".format(f, v)
Uniform LC:
  index_21 = 1.89689583705
  index_31 = 2.74650784403
\end{minted}

In fact, the same bug affects all of the features stored using a global
configuration: \texttt{Period\_fit}, \texttt{Psi\_CS}, \texttt{CAR\_tau}, \texttt{CAR\_mean}, 
the \textit{Fourier} components, and 
\texttt{StructureFunction\_index\_31} and \texttt{StructureFunction\_index\_32}, as mentioned above.

\item [Python exit]
	The Python language defines errors as 
    exceptions\footnote{Anomalous or exceptional conditions requiring special processing} 
    so in the presence of any misconfiguration, some exceptional states are created to 
    inform the caller code that "something has gone wrong."
	
    For example, if we want to write a division function that fails with a 
    divisor equal to 0, we could write the following.
\begin{minted}[]{pycon}
>>> def division(a, b):
...   if b == 0:
...	    raise Exception("b can't be 0")
...     return a / b
\end{minted}
The following result is obtained if the function is called.
\begin{minted}[]{pycon}
>>> result = division(1, 2.)
>>> print result
0.5
>>> result = division(1, 0)
Traceback (most recent call last):
  ...
  raise Exception("b can't be 0")
Exception: b can't be 0
\end{minted}

If it is desirable to print the null Python value \texttt{None} when
an exception occurs so it is possible to manage the error 
with the \texttt{try-except} construct.

\begin{minted}{pycon}
>>> try:
...   result = division(1, 0)
... except Exception:
...   result = None
>>> print result
None
\end{minted}

This simple example shows how Python can be used by the programmer
to correctly manage exceptional states: 
\textit{if you do not know how to deal with a configuration, then
throw an exception and ignore it without halting the system}.

The error exceptions provided to the programmer by Python 
can be reduced to two basic types: 
\texttt{BaseExceptions} and \texttt{Exceptions}. 
The first type comprises exceptions that may only be managed in very 
unusual cases, such as \texttt{SystemExit}.

The \texttt{SystemExit} exception errors are raised when the function 
\texttt{sys.exit()} is called. This function call
turns off the virtual machine and sends the exit code to the operating 
system. 
For example, the following piece of code:
\begin{minted}{pycon}
>>> import sys
>>> sys.exit()
\end{minted}
ends the Python virtual machine and sends a 0 value (no error) to the operating 
system.

In FATS, this occurs when a \texttt{FeatureSpace} 
is configured incorrectly and the Python virtual machine
is turned off. 
The system becomes unstable at 
least inside a multiprocessing environment (such as a web server, 
pipeline, or simple multi-core calculation). 
The following two codes reproduce this error.

\textbf{1- Ask for an invalid feature}
\begin{minted}{pycon}
$ python
Python 2.7.6 (default, ...) 
>>> import FATS
# ask for a nonexistent feature
>>> FATS.FeatureSpace(
...   featureList=['Foo'])
could not find feature Foo
# python ends here
\end{minted}

\textbf{2- Send an invalid configuration for a feature}
\begin{minted}{pycon}
$ python
Python 2.7.6 (default, ...) 
>>> import FATS
# ask for a nonexistent feature
>>> FATS.FeatureSpace(
...   featureList=['Std'], Std=(1,2,3))
error in feature Std
# python ends here
\end{minted}

The following code may manage this error.
\begin{minted}{pycon}
>>> import FATS
# ask for a nonexistent feature
>>> try:
>>>   FATS.FeatureSpace(
...     featureList=['Std'], Std=(1,2,3))
...  except:
...      # some manipulation
\end{minted}
However, as mentioned above, the \texttt{SystemExit} was not designed to be managed.

\item [Python 3]
Python 3 is a new language part of the Python family to 
replace the 2.7.x branch at 2020\footnote{\url{https://www.python.org/dev/peps/pep-0373/}}. 
This version is \textbf{backward incompatible} with Python 2.x 
but it includes several improvements in terms of expressibility and 
velocity\footnote{\url{https://speed.python.org/comparison/}}

Currently, all the foundations of the Python scientific\-stack, based on which FATS is built, 
have already been ported to the 3.x branch, but the code base of the project is still 
Python 2.x only\footnote{This issue are already been reported to the authors at: 
\url{https://github.com/isadoranun/FATS/issues/7}}. 
This issue essentially represents an impending death sentence for the package in the next two years.

\item [Light Curve Order]
This is a minor issue. Most of the light curve data sets represent the data 
in the following format: \textit{time/magnitude/magnitude-error}, whereas FATS uses the following format: 
\textit{magnitude/time/magnitude-error}. Thus, preprocessing is required in many cases. 

\item [Inefficient Routines]
In FATS, some performance issues are linked with the calculation of two features:
\texttt{MaxSlope} and \texttt{PeriodLS}. 
The code in the \texttt{MaxSlope} class is displayed below for analysis.

\begin{minted}[linenos]{python}
class MaxSlope(Base):
  """
  Examining successive (time-sorted) 
  magnitudes, the maximal first difference
  (value of delta magnitude over delta time)
  """
  def __init__(self):
    self.Data = ['magnitude', 'time']

  def fit(self, data):
    magnitude = data[0]
    time = data[1]
    slope = (
      np.abs(magnitude[1:] - magnitude[:-1]) / 
      (time[1:] - time[:-1]))
    np.max(slope)
    return np.max(slope)
\end{minted}
The \texttt{np.max} function is called two times in lines 16 and 17. 
The result from line 16 is not used, so this calculation can be removed.

The problem is more complicated for \texttt{PeriodLS} because the full 
implementation of the Lomb--Scargle Method \citep{vanderplas2018understanding}, 
which is included with FATS, is programmed in IDL Language \citep{landsman_idl_1995}
\footnote{\url{https://github.com/isadoranun/FATS/blob/master/FATS/lomb.py}} 
based on the published Numerical Recipes \citep{press_numerical_2007} routine.
This makes the code difficult to maintain and it has some performance issues 
due to the incorrect usage of the \textit{Numpy} library.  
Furthermore, this implementation of the Lomb--Scargle methodology is applied 
iteratively to calculate the nine Fourier features.

We extracted a high number of features for a particular light curve in the MACHO Survey 
and the computational time decreased by 20\% 
when all the features calculated using the Lomb--Scargle Periodogram
were neglected. 

\item [Testing and Coverage]
Measuring the qualitative and quantitative metrics for a software project involves
\textbf{unit testing} and \textit{code coverage}.

\textit{Unit testing} attempts to show that each part of the program is correct 
\citep{jazayeri_trends_2007} by isolating independent pieces of code and running tests on them.
Code coverage measures the percentage of code executed by the unit tests \citep{miller_systematic_1963}.

In the FATS tutorial, a static result is 
presented based on a test of invariance using 
unequal sampled data 
\footnote{\url{http://isadoranun.github.io/tsfeat/FeaturesDocumentation.html\#Appendix}}
The project has 19 automated unit testing cases and only one currently
fails\footnote{This issue has been reported to the authors 
at:\url{https://github.com/isadoranun/FATS/issues/9}}.
Unfortunately, the entire test suite only executes 62\% of the entire code, which is 
significantly below the requirement of 90\% adhered to by other astronomy projects such as 
Astropy \citep{robitaille_astropy:_2013}.

\item [Some features do not produce the expected values]
The FATS documentation \footnote{\url{http://isadoranun.github.io/tsfeat/FeaturesDocumentation.html}} states the following.
\begin{quote}
\begin{itemize}
  \item The feature \textit{StetsonK} for a Gaussian magnitude 
  distribution should take a value close to $2/\pi=0.798$.
  \item For a Gaussian magnitude distribution, \textit{StetsonJ} 
  should take a value close to zero.
  \item For a normal distribution the \textit{Anderson-Darling} 
  statistic should take values close to $0.25$.
\end{itemize}
\end{quote}

To validate the documentation, we calculated these three features for 
100,000 randomly generated Gaussian light curves and the results were 
not as expected. 
The results are presented in Table~\ref{tab:100kmontecarlo}, which
shows that the mean values for 
\textit{StetsonK},%$0.2\pm0.07$
\textit{AndersonDarling}, %$0.6052\pm0.26$, 
and \textit{StetsonJ} differ by several orders of magnitude from the expected values.
\begin{table}
\begin{tabular}{lrrr}
{} &  AndersonDarling &      StetsonJ &    StetsonK \\
\hline
count &           100000 &  1000000      &      100000 \\
mean  &           0.6052 &  5.953862e+05 &      0.2047 \\
std   &           0.2576 &  3.775754e+05 &      0.0662 \\
min   &           0.0930 &  2.442287e+05 &      0.0342 \\
25\%   &           0.3796 &  4.204386e+05 &      0.1564 \\
50\%   &           0.5993 &  5.104198e+05 &      0.2036 \\
75\%   &           0.8427 &  6.616722e+05 &      0.2510 \\
max   &           1.0000 &  4.056122e+07 &      0.4484 \\
\end{tabular}
\caption{Statistical analysis of the \textit{AndersonDarling}, \textit{StetsonJ}, and \textit{StetsonK} features executed with 100,000 randomly generated Gaussian light curves.}
\label{tab:100kmontecarlo}
\end{table}

\item [Missing Dependencies]

The current version of FATS (1.3.6) is distributed in the same manner as any standard Python third-party 
package via a service called "Python Package Index"
(PyPI)\footnote{\url{https://pypi.org/project/FATS/}}, 
which allows the package to be installed in Linux 
distributions with the following simple bash command.
\begin{minted}{bash}
$ pip install FATS
\end{minted}

However, attempting to run this command would lead to a "\texttt{missing packages}"
message and failure of the installation process. 
As mentioned above, FATS is built on top of the Python scientific-stack 
comprising libraries such as Numpy, Scipy, and Pandas, as well as specific libraries
including Matplotlib and StatsModels, which are required but not installed automatically.
The user is responsible for manually installing these packages in order to start 
working with the project. 
In addition, the standard Python scientific plotting library called Matplotlib is not used
for any of the core tasks in FATS, which is an unnecessary obstacle.

Most of these problems are fixed in the current code base in the GitHub repository, 
but no new releases have been made public since the aforementioned version 1.3.6 on June 7, 2015.

\item [Missing in-code documentation]
The only documentation for the project is the tutorial. Internally,
all of the components are undocumented. 

Two particular cases that are not mentioned in the tutorial are as follows.

\begin{itemize}
\item \textbf{Structure functions}: the \texttt{StructureFunction\_index\_21}, 
\texttt{StructureFunction\_index\_22}, and \texttt{StructureFunction\_index\_31} features
are part of the FATS codebase in GitHub (since February 9, 2016), 
but other than the comment posted in the version control commit stating that:
\textit{"Adding Structure Function from Simonetti et al. 1984"}(\footnote{Structure Functions commit: \url{https://github.com/isadoranun/FATS/commit/b45b5c1}} 
there are no descriptions of these features and their interpretations.

\item \textbf{Importing light curves toolbox}: The FATS documentation states the following. 

\begin{quotation}
  In addition to the features library, we provide a basic toolbox for importing and 
  preprocessing the data (\dots)

  (\dots) the function \texttt{ReadLC\_MACHO()} receives a MACHO id 
  (object id assigned in the MACHO survey) as an input and returns the 
  following output: magnitude measurement, time of 
  measurement, associated observational error (\dots)

  A demonstration of how to import a MACHO light-curve is presented below (\dots)
  \begin{minted}{python}
  lc_B = FATS.ReadLC_MACHO(
  	'lc_1.3444.614.B.mjd')  
  \end{minted}
\end{quotation}
However, the documentation is incorrect because \texttt{ReadLC\_MACHO()} receives a full path to a previously downloaded MACHO light curve.
\end{itemize}
\end{description}

% =============================================================================
% SECTION REENGENEERING
% =============================================================================
\section{Code Refactoring: Improving FATS }
\label{section:reengineering}
In Section~\ref{section:intro}, we explained that bad 
design choices can be addressed using the code refactoring technique, thereby
modifying the design of FATS but without changing its functionality.
This is ensured by the following method.
\begin{enumerate}
\item First, all of the feature extractors in FATS 
	are executed for an example light curve and the results are stored.
\item Second, a unit test case is conducted, which executes the same extractors and checks
	whether the results are the same as those stored in the stored file.
\item Next, porting of the new architecture is commenced progressively while 
	checking whether the test is still passed.
\item If an extractor directly changes the result 
	for any feature extractor, then the test is modified in order to assess this change.
\end{enumerate}
This approach where a test is specified and the code is then built is 
called test-driven development. 
This technique makes it easier to trust that each new piece of code written 
during the refactoring process is not generating new regression bugs\footnote{Regression is something that used to work, but no longer does}.

After the initial test, the code refactoring process is divided into the following six
sequential tasks.
\begin{enumerate}
\item Update each feature extraction function while 
	maintaining the same behavior\footnote{the results 
    must be the same for the same input in order to pass the 
    initial test}.
\item Write a new \texttt{FeatureSpace} stateless class. 
\item Test to ensure that each feature extractor returns reasonable values, at
	least for usual expected inputs.
\item Continue by performing more tests until $90\%$ code coverage is achieved.
\item Include documentation for the features and extractors 
	inside the code as Python 
    docstrings\footnote{\url{https://docs.scipy.org/doc/numpy-1.14.0/reference/}}.
\item Port the tutorial. This process includes auto-generating
	documentation for every feature extractor.
\end{enumerate}

% =============================================================================
% SECTION RESULT
% =============================================================================
\section{Results: {feATURES eXTRACTOR for tIME sERIES} (\textit{feets})}
\label{section:results}
Due to the radical restructuring of the project,
we decided to create a new package called 
\textit{\textbf{feATURE eXTRACTOR FOR tIME sERIES}} (\textit{feets}).
The FATS functionalities can be found inside this new \textit{feets} library 
and most of the issues described in Section~\ref{section:criticism} have been addressed. 
Error handling strategies were applied in cases where this was not possible
in order to inform the user about any possible issues with the results.

\subsection{Clear code, high coverage rate, and public API documentation}
Testing in \textit{feets} includes 49 unit test cases that are currently passed,
with up to $90\%$ code coverage. 
The source code is completely PEP8 style compliant. 
A continuous integration\footnote{\url{https://travis-ci.org/carpyncho/feets}} 
tool automatically executes the tests whenever a new version of the project
is uploaded to the public code repository.
Finally, documentation was written for most of the public objects, 
classes, and functions in \textit{feets}.
This documentation is compiled automatically and published on the 
following page: \url{http://feets.readthedocs.io}.
\subsection{Support for Python 3.x}
The current version of \textit{feets}\footnote{\url{https://pypi.org/project/feets/\#history}}
(0.4) is compatible with Python versions 2.7, 3.5, 3.6, and 3.7, where
compatibility is achieved via the \textit{six} 
library\footnote{\url{https://pypi.org/project/six/}}.

\subsection{Better encapsulation of extractors}
The extractors were redesigned in order to avoid global variables 
and they now return a fixed set of features because the \texttt{FeatureSpace}
is in charge of discarding the unused ones.
The new infrastructure is capable of making compile-time checks for 
any extractor introduced by a user in order to 
avoid unexpected behavior during feature extraction.

Finally, a \texttt{feets.register\_extractor} function is provided, which os
capable of including user-defined extractor classes in the \textit{feets} functionalities.
More detailed explanations of these topics can be found in the tutorial 
(\url{http://feets.readthedocs.io/en/latest/#Library-structure}).
\subsection{Exceptions and Warnings}
If the user misconfigures the \texttt{feets.FeatureSpace}, 
an exception is raised instead of \texttt{sys.exit()}, which shuts down 
the entire Python virtual machine. 
In addition, a warning is shown when the user requests a feature from any of the 
extractors with inconsistent behavior: (\textit{StetsonK}, 
\textit{StetsonJ} \citep{richards_machine-learned_2011}, and \textit{Anderson-Darling} \citep{kim_-trending_2009}).	

\subsection{Integration with Astropy and other dependencies}
The homepage of the Astropy\footnote{\url{http://www.astropy.org}} 
project \citep{robitaille_astropy:_2013} states the following.
\begin{quotation}
The Astropy Project is a community effort to develop a core package for astronomy using the Python 
programming language and improve usability, interoperability, and collaboration between astronomy 
Python packages. 
\end{quotation}
Our decision to include the Astropy package 
was dependent on two objectives: to replace the built-in 
Lomb-Scargle
\footnote{\url{https://github.com/isadoranun/FATS/blob/6fcd852adf213a477fda878650be5b467a5dd0d8/FATS/lomb.py}} 
implementation of FATS with
that distributed by the Astropy project in order to improve the performance of the
period feature extractor; and to propose \textit{feets} as part of the 
Astropy-Afiliated-Packages\footnote{\url{http://www.astropy.org/affiliated/index.html}}
in order to demonstrate a commitment to Astropy's goals for Python astronomy and astrophysics packages.

In addition to Astropy, each dependency in \textit{feets} 
is included into the Python-Package-Index installer, so the project is ready to use
with a single \texttt{pip install feets} command.
Finally, Matplotlib was removed as a dependency.
\subsection{Other Enhancements}
\begin{itemize}
\item The order of the input parameters was updated to 
	\textit{Time-Magnitude-Magnitude Error}, which is now
    consistent with most previous studies.
\item The preprocessing functions were renamed in order to make 
	them more intuitive. For example,
\begin{minted}[]{python}
FATS.Preprocess_LC(
  mag, time, error).Preprocess()
FATS.Align_LC(
  time, time2, mag, mag2, error, error2)
\end{minted}
was renamed as follows.
\begin{minted}[]{python}
feets.preprocess.remove_noise(
  time, mag, error)
feets.preprocess.align(
  time, time2, mag, mag2, error, error2)
\end{minted}

\item The \texttt{feets.dataset} module was included to retrieve
	light curves from the MACHO, OGLE-III\citep{udalski_optical_2004} 
    survey as well as for creating synthetic light curves based on a set of random distributions of parameters (e.g., periods).
\end{itemize}

% =============================================================================
% SECTION CONCLUSIONS
% =============================================================================
\section{Conclusions and Future Work}
\label{section:conclusions}
In this study, we redesigned a Python library for extracting times series 
features, where we addressed the design 
flaws of its predecessor. 
This new \textit{feets} package is compatible with upcoming Python versions and it is fully documented.
This package includes a larger test suite with $90\%$ code coverage,
greater extensibility, integration with more data sets, 
and has been being proposed to be including as Astropy affiliated package.
The project was developed in a public repository,
with more than 233 commits and five contributors. 
Future development will focus on the following three areas: incorporating
new features, improving documentation, and 
possible integration with tools for interactively analyzing features.

% =============================================================================
% SECTION Acknowledgments
% =============================================================================
\section{Acknowledgments}
The authors would like to thank their families and friends, as well as the IATE
astronomers for useful comments and suggestions.

This work was partially supported by the Consejo Nacional
de Investigaciones Cient\'ificas y T\'ecnicas (CONICET, Argentina)
and the Secretaría de Ciencia y Tecnolog\'ia de la Universidad
Nacional de C\'ordoba (SeCyT-UNC, Argentina).
J.B.C, F.R., and B.S. were supported by a fellowship from CONICET.
Some processing was achieved with the Argentine VO (NOVA) infrastructure,
for which the authors express their gratitude.

This research employed the
http://adsabs.harvard.edu/, Cornell University xxx.arxiv.org repository,
the Python programming language, the Numpy and Scipy libraries,
and the other packages utilized can be found at the 
GitHub webpage for feets.

% =============================================================================
% SECTION BIBLIO
% =============================================================================
%
\section*{References}
\label{biblio}
\bibliographystyle{model2-names-astronomy}
\bibliography{feets}

\end{document}